\documentclass[final,5p,times,twocolumn]{elsarticle}

\usepackage[T1]{fontenc}
\usepackage[utf8]{inputenc}

\usepackage{amsmath}
\usepackage{amsfonts}
\usepackage{amssymb}
\usepackage{amsxtra}
\usepackage{array}
\usepackage{color}
\usepackage{dcolumn}
\usepackage{graphicx}
\usepackage{hepunits}
\usepackage{xspace}
\usepackage{bm}

\definecolor{purple}{rgb}{0.5,0,0.5}
\definecolor{blue}{rgb}{0.0,0,0.9}
\definecolor{prdblue}{rgb}{0.133,0.118,0.498}
\usepackage[colorlinks=true, pdfstartview=FitV, linkcolor=prdblue, citecolor= prdblue, urlcolor=prdblue]{hyperref}

\usepackage[mathscr,scaled=1.15]{urwchancal}
\DeclareFontFamily{OT1}{pzc}{}
\DeclareFontShape{OT1}{pzc}{m}{it}
{<-> s * [1.15] pzcmi7t}{}
\DeclareMathAlphabet{\mathpzc}{OT1}{pzc}{m}{it}

\biboptions{sort&compress}

\journal{Physics Letters B}

\makeatletter

\setbox0\hbox{$\xdef\scriptratio{\strip@pt\dimexpr
    \numexpr(\sf@size*65536)/\f@size sp}$}

\newcommand{\scriptveryshortarrow}[1][3pt]{{%
    \hbox{\rule[\scriptratio\dimexpr\fontdimen22\textfont2-.2pt\relax]
               {\scriptratio\dimexpr#1\relax}{\scriptratio\dimexpr.4pt\relax}}%
   \mkern-4mu\hbox{\let\f@size\sf@size\usefont{U}{lasy}{m}{n}\symbol{41}}}}

\makeatother

\begin{document}
%\begin{CJK}{UTF8}{song}

\begin{frontmatter}

\title{Imaging the charge distributions of flavor-symmetric and -asymmetric mesons}

\author[UHe,UPO]{Yin-Zhen~Xu % yinzhen.xu@dci.uhu.es
       $^{\href{https://orcid.org/0000-0003-1623-3004}{\textcolor[rgb]{0.00,1.00,0.00}{\sf ID}},}$}
       
\author[UHe]{Adnan Bashir%
       $^{\href{https://orcid.org/0000-0003-3183-7316}{\textcolor[rgb]{0.00,1.00,0.00}{\sf ID}},}$}       

\author[UHe]{Kh\'epani Raya%
    $^{\href{https://orcid.org/0000-0001-8225-5821}{\textcolor[rgb]{0.00,1.00,0.00}{\sf ID}},}$}
%\email[]{khepani.raya@dci.uhu.es}

\author[UHe]{Jos\'e Rodr\'{\i}guez-Quintero%
       $^{\href{https://orcid.org/0000-0002-1651-5717}{\textcolor[rgb]{0.00,1.00,0.00}{\sf ID}},}$}
%\author[UHe]{Jos\'e Rodr\'{\i}guez-Quintero}
%\ead{jose.rodriguez@dfaie.uhu.es}

\author[UPO]{Jorge Segovia
       $^{\href{https://orcid.org/0000-0001-5838-7103}{\textcolor[rgb]{0.00,1.00,0.00}{\sf ID}},}$}

\address[UHe]{
Dpto.\ Ciencias Integradas, Centro de Estudios Avanzados en Fis., Mat. y Comp., Fac.\ Ciencias Experimentales, Universidad de Huelva, Huelva 21071, Spain}
\address[UPO]{Dpto. Sistemas F\'isicos, Qu\'imicos y Naturales, Univ.\ Pablo de Olavide, E-41013 Sevilla, Spain\\[1ex]
%
%\hspace*{-8em}Email addresses:
\href{mailto:yinzhen.xu@dci.uhu.es}{yinzhen.xu@dci.uhu.es} (Yin-Zhen Xu);
\href{mailto:adnan.bashir@dfa.uhu.es}{adnan.bashir@dfa.uhu.es} (Adnan Bashir)
\href{mailto:khepani.raya@dci.uhu.es}{khepani.raya@dci.uhu.es} (Kh\'epani Raya);
\href{mailto:jose.rodriguez@dfaie.uhu.es}{jose.rodriguez@dfaie.uhu.es} (Jos\'e Rodr{\'{\i}}guez-Quintero);
\href{mailto:jsegovia@upo.es}{jsegovia@upo.es} (Jorge Segovia)
%
%\\[1ex]
%%%
%Date: 2025 Nov 15\\[-6ex]
%%% Begin date 2023 Sep 27
}

\begin{abstract}
We investigate the internal structure of a comprehensive set of pseudoscalar and vector mesons, including both flavor-symmetric and flavor-asymmetric systems, by reconstructing their charge distributions from electromagnetic form factors. To achieve this, we employ a Maximum Entropy Method optimized for charge distributions, utilizing previously published form factor data obtained within the Dyson–Schwingers and Bethe–Salpeter equations framework.  Furthermore, we calculate the average distance between the valence quark and antiquark that constitute the meson, interpreting it as an estimate for both the meson’s spatial size and the typical range of quark motion. Our results reveal that this distance for the lightest quarkonia is approximately five times larger than that for the heaviest. Moreover, due to spin effects, vector mesons exhibit sizes that are 5–15\% larger than their pseudoscalar counterparts.
\end{abstract}

\begin{keyword}
charge distribution \sep
heavy-light meson \sep
Dyson-Schwinger/Bethe-Salpeter equations \sep
electromagnetic form factors 
\end{keyword}

\end{frontmatter}
%\end{CJK}

%%%%%%%%%%%%%%%%%%%%%%%%%%%%%%%%%%%%%%%%%%%%%%%%%%%%%%%%%%%%%%%%%%%%%%%%%%%%%%%%%%%%%%%%%%%%%%%%%%%%%%%%%%%%%%%%%%%%%%%
% 4500 words

\section{Introduction}
%\noindent\emph{1.$\;$Introduction}.
%

The study of hadron properties remains a central topic in particle physics. Among these properties, the charge distribution (namely, the spatial distribution of electric charge within a hadron) offers critical insight into the internal structure of strongly interacting particles and has attracted significant attention \cite{Miller:2007uy, Epelbaum:2022fjc, Kim:2022bia, Miller:2010tz,Lorce:2020onh, Freese:2021mzg, Chen:2022smg, Xu:2024vkn}. In particular, the spatial distribution of positive (or negative) charge can be directly linked to the static arrangement of valence (anti)quarks, providing valuable insight into Quantum Chromodynamics (QCD) bound states as probed by electromagnetic interactions.

Among hadrons, flavor-asymmetric mesons comprising one heavy and one light quark are especially important. These systems uniquely incorporate both relativistic and non-relativistic dynamics, owing to the large mass disparity between the constituent quarks. This asymmetry gives rise to distinctive structural and dynamical properties, making such mesons ideal laboratories for investigating strong interactions and the quark confinement mechanism \cite{LHCb:2021vsc, BaBar:2021ich, Moita:2021xcd}. Analyzing their charge distributions would enable us to understand the long- and short-distance behavior of the color force.

The charge distribution is traditionally derived from the electromagnetic form factor via a three-dimensional (3D) Fourier transform \cite{RevModPhys.30.482, Ernst:1960zza, Sachs:1962zzc}. However, this method has been recently questioned. Several studies suggest that a two-dimensional (2D) Fourier transform, or even more advanced frameworks, may be more suitable for relativistic systems \cite{Miller:2007uy, Epelbaum:2022fjc, Kim:2022bia,Miller:2010tz, Miller:2018ybm, Jaffe:2020ebz,Freese:2023abr}. Regardless of the dimensionality, such transforms require complete data in the space-like region of the form factor. In practice, this data is limited, making the reconstruction of a charge distribution --usually a mathematically ill-posed problem-- a considerable numerical challenge. In Ref.\,\cite{Xu:2024vkn}, we developed a computational framework based on the Maximum Entropy Method (MEM) to address this issue and successfully applied it to light mesons and quarkonia. 

A suitable and systematically improvable theoretical framework for calculating electromagnetic form factors of hadrons is provided by the coupled formalism of Dyson–Schwinger and Bethe–Salpeter equations (DSEs/BSEs) of QCD. This non-perturbative, Poincaré-covariant approach enables the simultaneous treatment of confinement and dynamical chiral symmetry breaking, while smoothly recovering the perturbative regime in the weak-coupling limit. It has been successfully applied to the study of hadron properties for over three decades \cite{Roberts:1994dr, Eichmann:2016yit, Raya:2024ejx, Maris:1999nt, Maris:1997tm, Qin:2011dd, Rodriguez-Quintero:2010qad, Qin:2011xq, Xu:2019ilh, Xu:2020loz, Qin:2019oar, Xu:2021lxa, Li:2023zag, Xu:2023izo,Xu:2025hjf,daSilveira:2022pte}.

In contrast to quarkonium systems, open-flavor mesons pose additional challenges due to flavor asymmetry, most notably in the formulation of physically meaningful and mathematically consistent interaction kernels. Several approaches have been developed to address this issue, each employing different interaction kernels \cite{Hilger:2017jti,Chen:2019otg, Serna:2020txe, Serna:2022yfp, daSilveira:2022pte, Serna:2024vpn, Xu:2025cyj, Qin:2020jig, Xu:2022kng,Gao:2024gdj}. Among them, particularly promising is the framework put forward in Refs.\,\cite{Qin:2020jig, Xu:2022kng}, in which all Ward–Green–Takahashi identities (WGTIs) are rigorously preserved. However, a major limitation arises from the analytic structure of quark propagators in the complex plane, which restricts calculations to a finite, low-momentum transfer region \cite{Sanchis-Alepuz:2017jjd, Windisch:2016iud, Chen:2018rwz, Fischer:2005en}. Consequently, electromagnetic form factor across the full kinematic domain is inaccessible. This makes the use of reconstruction techniques, such as MEM, essential for obtaining physically meaningful charge distributions from incomplete data.

Electromagnetic form factors for a comprehensive set of vector and pseudoscalar mesons, including both flavor-symmetric and flavor-asymmetric systems, have been computed within the DSEs/BSEs framework using an effective interaction kernel in which the axial-vector WGTIs are approximately preserved \cite{Xu:2024fun,Xu:2025hjf,Xu:2024rew}. In the present work, we build upon these results to extend the implementation of MEM, which was previously applied to, basically, flavor-symmetric mesons \cite{Xu:2024vkn}, to reconstruct the charge distributions of all ground-state vector and pseudoscalar mesons. In addition to presenting the reconstructed charge distributions, we analyze the static spatial distribution of valence quarks. We report results both in 3D and 2D transverse space to ensure the robustness of the entailed implications.

\section{Reconstructing charge distribution from electromagnetic form factors}
\label{sec:2}

It is well known that three- and two-dimensional charge distributions can be derived from the electric form factors using the corresponding Fourier transformation \cite{RevModPhys.30.482, Ernst:1960zza, Sachs:1962zzc} 
\begin{subequations}
\begin{align}
\label{eq:3d}
F(Q^2)&=4\pi\int_0^{\infty} dr\, j_0(Qr) r^2 \rho^{\rm{3D}}(r) \\
&=2\pi \int_0^{\infty} db\, J_0(Qb) b \rho^{\rm{2D}}(b) \,,
\label{eq:2d}
\end{align}
\end{subequations}
where $F(Q^2)$ stands for the electric form factor for a given meson, $j_0$ ($J_0$) for the zeroth spherical (standard first-kind) Bessel function, and $\rho^{\rm{3D}}(r)$ or $\rho^{\rm{2D}}(b)$ denote the charge distribution resulting from the usual 3D or the transverse 2D non-relativistic approach. As discussed in the introduction, some controversy \cite{Jaffe:2020ebz,Freese:2021mzg,Epelbaum:2022fjc,Xu:2023izo} has recently arisen regarding the interpretation and reliability of non-relativistic objects as 3D or transverse 2D densities, the latter being defined for the charge as \cite{Miller:2010tz, Miller:2018ybm, Burkardt:2002hr}. In the vector case, we use $F(Q^2) = G_E(Q^2)$ and ignore corrections stemming from projecting form factors defined in 3D to 2D \cite{Kim:2022bia}. An exact relation between charge distribution and form factor remains under investigation \cite{Jaffe:2020ebz,Li:2022hyf}.

If the form factor data are known across the entire space-like region, the charge distributions can be derived by inverting the Fourier transform (FT) in Eqs.~\eqref{eq:3d} and \eqref{eq:2d}. However, accessing the large momentum domain remains usually challenging and only the low momentum region of $F(Q^2)$ can be exploited. Alternatively, one might rely on dispersion relations to map the time-like data on the form factor onto the space-like domain, although this approach is sensitive to subtleties in the asymptotic behavior of the form factor\,\cite{RuizArriola:2025wyq}. Therefore, a procedure other than the pure FT inversion needs to be implemented to deal with truncated data for the form factors in order to extract charge distributions. 

In Ref.\,\cite{Xu:2024vkn}, we addressed this problem further, suggesting a method which was then thoroughly discussed and illustrated through the explicit calculation of charge distributions of mostly flavor-symmetric pseudoscalar and vector mesons. The main idea is that both Eqs.\,\eqref{eq:3d} and \eqref{eq:2d} can be seen as particular cases of  
\begin{equation}\label{eq:FTgen}
F(Q^2)=\int_{0}^{\infty} d\tilde{r}\, \mathcal{K}(Q,\tilde{r})\rho(\tilde{r}) \,,
\end{equation}
where $\mathcal{K}(Q,\tilde{r})$ is the kernel of integration, either $4\pi {\tilde{r}}^2 j_0(Q\tilde{r})$ or $2\pi \tilde{r} J_0(Q\tilde{r})$, with $\tilde{r}=r$ or $b$, respectively, for 3D and 2D. The determination of the charge profile $\rho(\tilde{r})$, when $F(Q^2)$ is only partially known, might be considered a mathematically ill-posed problem. However, driven by recent progress in computational power, a wide variety of modern numerical methods have been developed to take up this challenging task. Among these methods, the MEM has recently been applied successfully to analogous problems in hadron physics \cite{Asakawa:2000tr, Nickel:2006mm, Qin:2014dqa, Gao:2016jka, Xu:2021lxa, Mueller:2010ah, Zhang:2023oja}. 

Following MEM approach, $\rho(\tilde{r})$ is reconstructed by maximizing the following functional 
\begin{equation}
\label{eq:MEM}
\mathcal{Q}[\rho]=\alpha S[\rho]-L[\rho],
\end{equation}
where $L[\rho]$ is the likelihood function 
\begin{equation}
\label{eq:Lrho}
L[\rho] = \sum_i \frac{1}{2\sigma^2_i} \left[F(Q_i^2) - \int_0^\infty d\tilde{r}\, \mathcal{K}(Q_i,\tilde{r}) \rho(\tilde{r}) \right]^2,
\end{equation}
which corresponds to standard $\chi^2$-fitting, where $\sigma_i$ denotes the error in the $i$-th data point, $\alpha$ is the regularization parameter and $S[\rho]$ represents the Shannon-Jaynes entropy: 
\begin{equation}
\label{eq:entropy}
S\left[\rho\right] = \int_0^{\infty} d\tilde{r}\, \left[\rho(\tilde{r})-\rho_0(\tilde{r})-\rho(\tilde{r}) \log \left(\frac{\rho(\tilde{r})}{\rho_0(\tilde{r})}\right)\right],
\end{equation}
where $\rho_0(\tilde{r})$ denotes a prior estimate for $\rho(\tilde{r})$. The impact on the charge profile from the choice of the prior estimate has been studied at length in Ref.\,\cite{Xu:2024vkn}. It is inferred that the reliable profiles can be extracted from considering the well known vector-meson-dominance (VMD) model. In practical calculations, the integral Eqs.~(\ref{eq:MEM}-\ref{eq:entropy}) require to be discretized into matrix equations and the Singular Value Decomposition is applied for dimensional reduction. More technical details about MEM can be found in Refs.\,\cite{Asakawa:2000tr,Rothkopf:2011ef,Skilling1989}.

In this work, as stated in the introduction, the results of Ref.\,\cite{Xu:2024vkn} are extended to all ground-state pseudoscalar and vector mesons, specially those exhibiting a strong flavor asymmetry. To achieve this, the input data for Eq.\,\eqref{eq:Lrho} are taken from Ref.\,\cite{Xu:2024fun}, where the electromagnetic form factors for the mesons, and their separate contributions from quark and antiquark were calculated within the DSEs/BSEs framework in a momentum range restricted up to $Q^2 \sim 2$ GeV$^2$ (see Figs.\,4 and 5 therein). In the standard MEM~\cite{Asakawa:2000tr}, the best $\alpha$ and $\rho(\tilde{r})$ are determined in terms of probability as long as $\sigma^2_i$ are properly known. Such an evaluation of the uncertainties in the numerical computations within the DSEs/BSEs framework is a challenging task. Alternatively, a cross-validation method, widely adopted in machine learning, is applied here \cite{Xu:2024vkn}. First, the errors in the data for the form factors are assumed to be all the same, $\sigma_i=\sigma$\, \cite{Mueller:2010ah}. Then, basically, the ensemble of data is divided into two sub-sets, namely a training one containing about $70 \%$ of data, and a testing one with the remaining $30 \%$. A value for $\bar{\alpha}=\alpha/\sigma^2$ and the corresponding distribution function $\rho(\tilde{r})$ are determined with the trained sub-set, and the optimal $\bar{\alpha}$ is selected based on its predictive performance on the testing sub-set. The same is repeated by considering many different ways of choosing training and testing sub-sets. The final profile for the charge distribution is obtained by averaging all the different outputs, and the statistical error is correspondingly estimated through quadratic dispersion. The numerical results show that the total uncertainties are dominated by the effect of the sensitivity to the prior estimates, assessed by rescaling the VMD model by a factor 0.5 to 2. An error band is then defined by merging the statistical errors with these systematic effects.

\section{Numerical outputs}
\label{sec:3}

\begin{figure*}[htbp]
\centering 
\includegraphics[width=.95\textwidth]{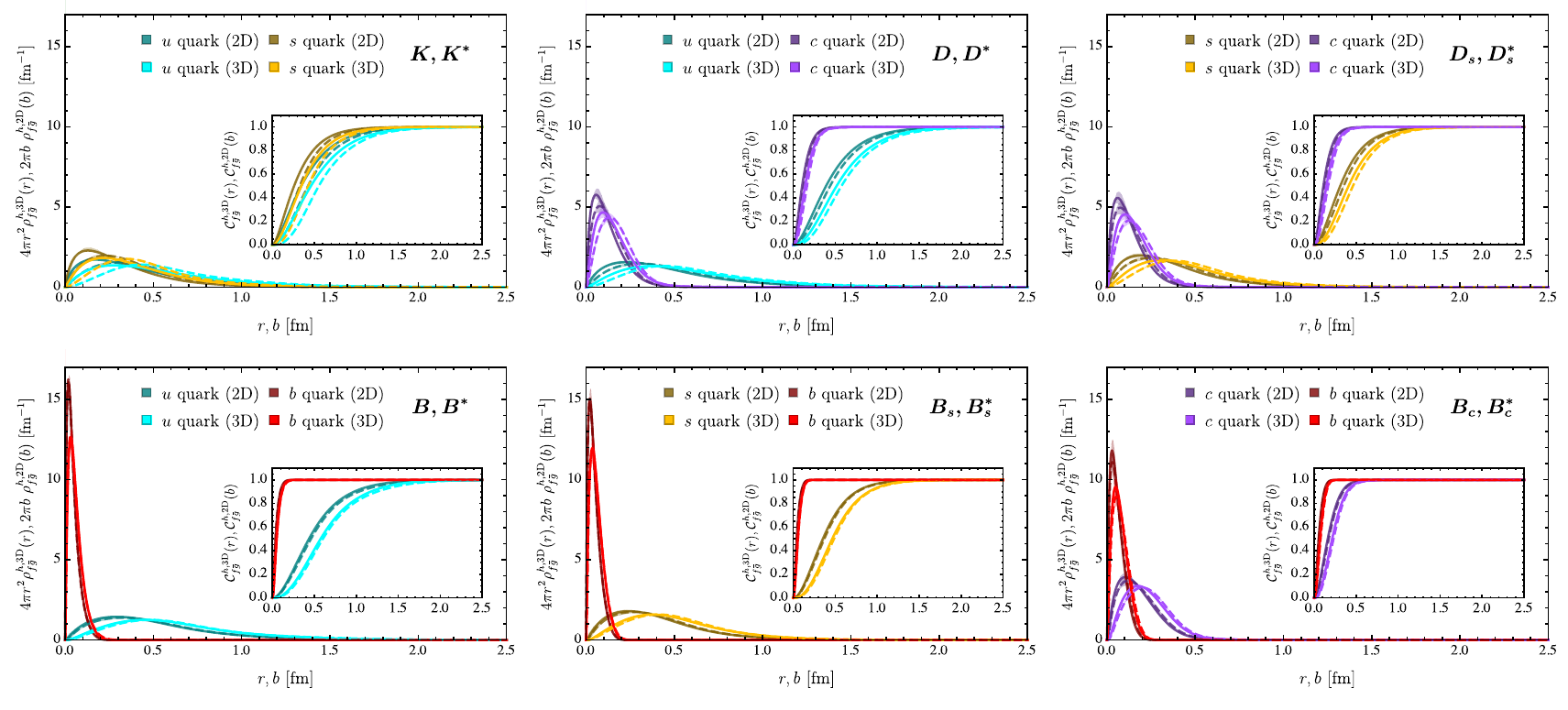}
\caption{\label{fig:quark_distribution} 
In flavor-asymmetric mesons, the static spatial distribution of valence (anti)quarks (see, Eq.\,\eqref{eq.charge}) and corresponding cumulative distribution function (see, Eq.\,\eqref{eq.CDF}). Solid lines: pseudoscalar mesons; dashed lines: vector mesons.}
\end{figure*}

\begin{figure*}[htbp]
\centering 
\includegraphics[width=.99\textwidth]{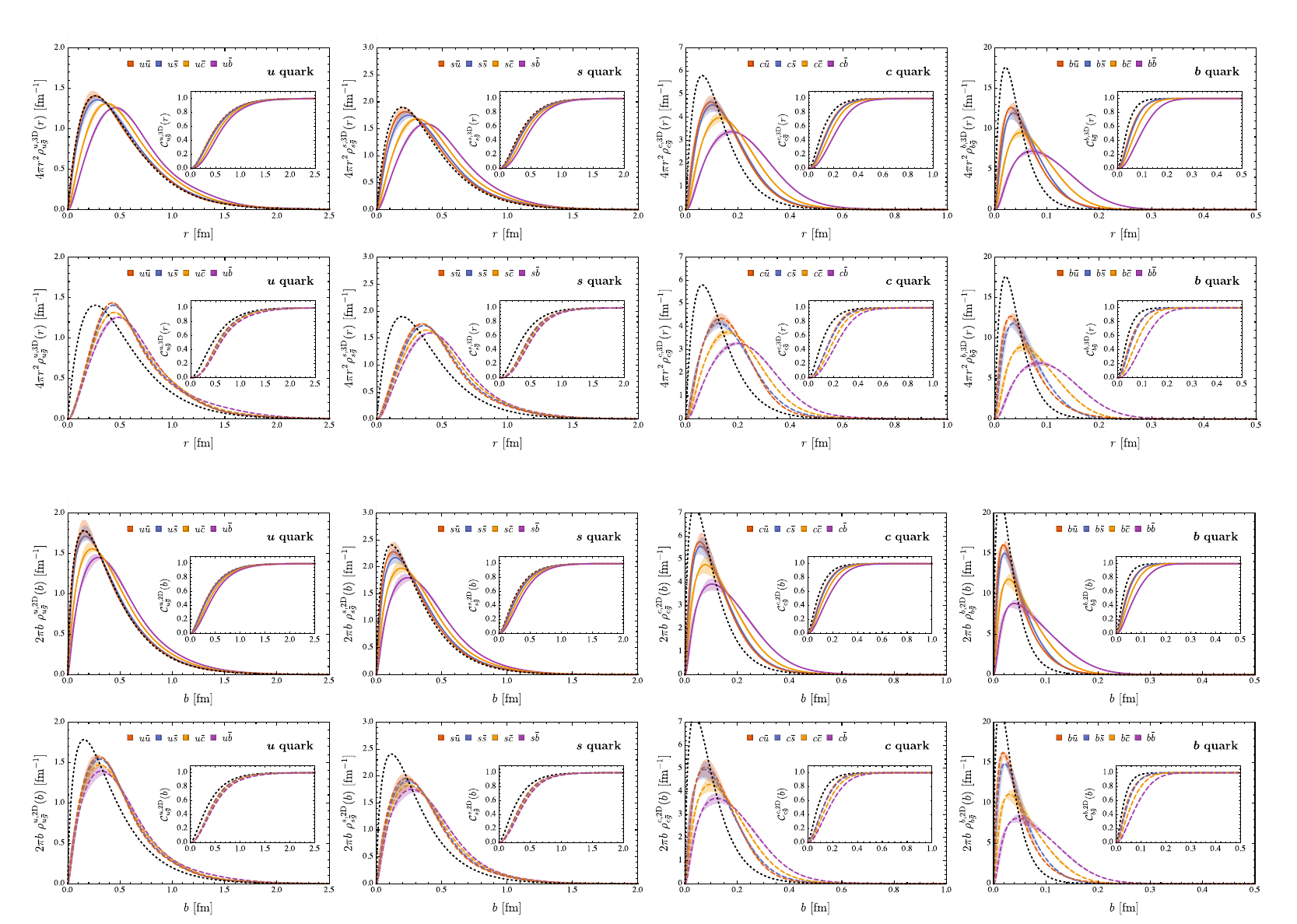}
\caption{\label{fig:quark} The static spatial distribution of quarks with the same flavors in different heavy-light mesons, and corresponding cumulative distribution function. Solid lines: pseudoscalar mesons; dashed lines: vector mesons; black dotted lines: VMD. The first and second rows are the 3D case, and the third and fourth rows are the 2D case.}
\end{figure*}

\begin{figure*}[htbp]
\centering 
\includegraphics[width=0.95\textwidth]{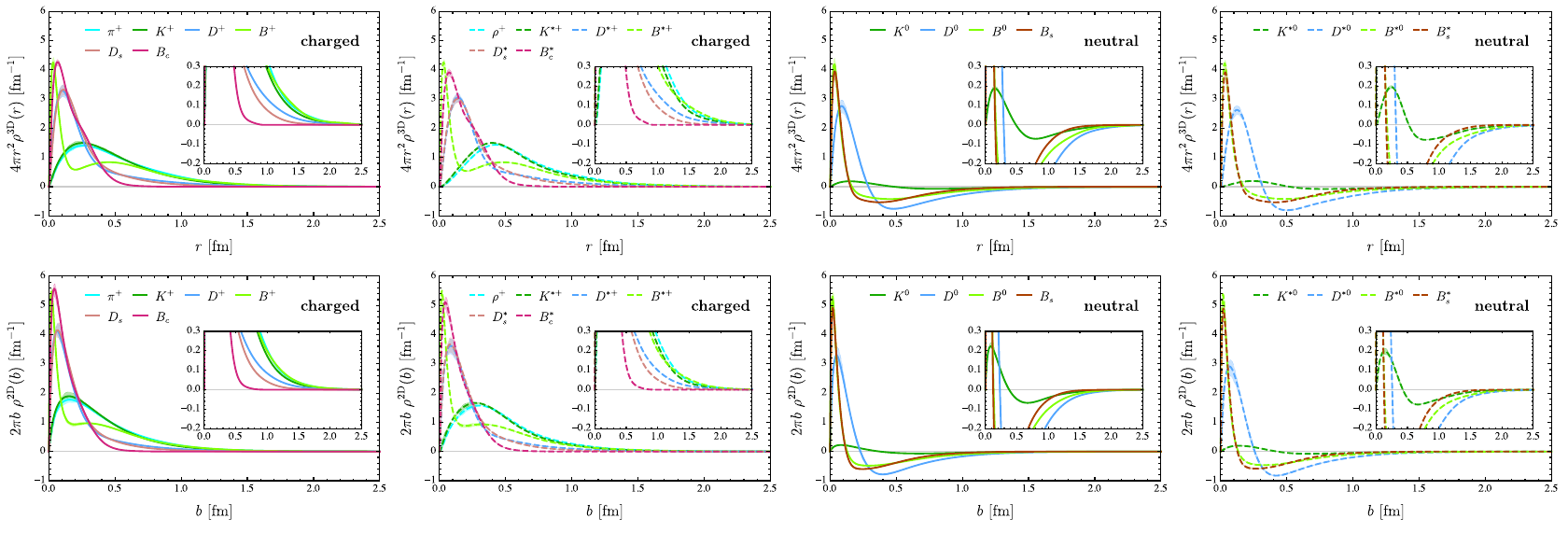}
\caption{\label{fig:meson} The charge distribution of pseudoscalar/vector flavor-asymmetric mesons. The first row stands for 3D and the second for 2D.}
\end{figure*}

Following the procedure outlined in the previous section, the overall charge distribution for an $f\bar{g}$ meson can be written as
  \begin{equation}
  \rho^{3 \mathrm{D}}_{f\bar{g}}(\tilde{r}) = \hat{Q}^f  \rho^{f,3 \mathrm{D}}_{f\bar{g}}(\tilde{r}) + \hat{Q}^{\bar{g}}  \rho^{\bar{g},3 \mathrm{D}}_{f\bar{g}}(\tilde{r})\,.
  \label{eq.charge}
\end{equation}
That is, it is the superposition of the charge density profiles, $\rho^{h,3 \mathrm{D}}_{f\bar{g}}(\tilde{r})$, reconstructed from $F^{h}_{f\bar{g}}(Q^2)$, contribution of quark and antiquark flavor $h=f,\bar{g}$ to the meson's total form factor \cite{Bhagwat:2006pu,Xu:2024fun}. An analogous expression can be written down for the transverse (2D) charge densities. 
$\rho^{h,3 \mathrm{D}}_{f\bar{g}}(\tilde{r})$ can be then interpreted as the static spatial distribution of the valence (anti)quarks under electromagnetic probe, which is positive definite\footnote{It should be noted that the reconstruction method outlined in Sec.\,\ref{sec:2} is only applicable to extract positive definite distributions.} and allows for the definition of the following cumulative distribution functions
\begin{equation}
\hspace{-2mm} \mathcal{C}^{h,3 \mathrm{D}}_{f\bar{g}}\left(r^\prime\right)= \hspace{-2mm} \int_0^{r^\prime} \hspace{-3mm} d r \, 4 \pi r^2 \rho_{f\bar{g}}^{h,3 \mathrm{D}}(r), \;\mathcal{C}^{h,2 \mathrm{D}}_{f\bar{g}}\left(b^\prime\right)= \hspace{-2mm} \int_0^{b^\prime} \hspace{-3mm} d b \, 2 \pi b \rho_{f\bar{g}}^{h,2 \mathrm{D}}(b) \,;
 \label{eq.CDF}
\end{equation}
where $ \mathcal{C}^{h,3 \mathrm{D}}_{f\bar{g}}\left(\infty\right)= \mathcal{C}^{h,2 \mathrm{D}}_{f\bar{g}}\left(\infty\right)=1$ owing to the normalization of the spatial distributions.

\begin{table}[htbp]
\centering
\begin{tabular}{l c c | l c c}
\hline
%\multicolumn{6}{c}{3D} \\ \hline
& Eq.~(\ref{eq.slope}) & Eq.\,(\ref{eq.int}) & & Eq.~(\ref{eq.slope}) & Eq.\,(\ref{eq.int}) \\ \hline
\multicolumn{6}{c}{3D} \\ \hline
$\pi$       & 0.646     & 0.646(1)     & $\rho$       & 0.722     & 0.722(1)     \\
$K$         & 0.608     & 0.608(1)     & $K^{*+}$     & 0.679     & 0.679(1)     \\
$D^+$       & 0.435     & 0.435(1)     & $D^{*+}$     & 0.473     & 0.473(1)     \\
$B^+$       & 0.619     & 0.620(1)     & $B^{*+}$     & 0.657     & 0.656(1)     \\
$D_s$       & 0.352     & 0.352(1)     & $D^*_s$      & 0.381     & 0.381(1)     \\
$B_c$       & 0.219     & 0.219(1)     & $B^*_c$      & 0.231     & 0.231(1)     \\ \hline
$K^0$       & 0.253$i$  & 0.253(1)$i$  & $K^{*0}$     & 0.274$i$  & 0.274(1)$i$  \\
$D^0$       & 0.556$i$  & 0.555(1)$i$  & $D^{*0}$     & 0.604$i$  & 0.604(1)$i$  \\
$B^0$       & 0.435$i$  & 0.436(1)$i$  & $B^{*0}$     & 0.462$i$  & 0.463(1)$i$  \\
$B_s$       & 0.337$i$  & 0.337(1)$i$  & $B^*_s$      & 0.347$i$  & 0.347(1)$i$  \\ \hline
\multicolumn{6}{c}{2D} \\ \hline
%& Slope & Integration & & Slope & Integration \\ \hline
$\pi$       & 0.527     & 0.527(1)     & $\rho$       & 0.590     & 0.590(1)     \\
$K$         & 0.496     & 0.497(1)     & $K^{*+}$     & 0.554     & 0.554(1)     \\
$D^+$       & 0.355     & 0.355(1)     & $D^{*+}$     & 0.386     & 0.387(1)     \\
$B^+$       & 0.505     & 0.507(1)     & $B^{*+}$     & 0.536     & 0.537(1)     \\
$D_s$       & 0.287     & 0.288(1)     & $D^*_s$      & 0.311     & 0.311(1)     \\
$B_c$       & 0.179     & 0.179(1)     & $B^*_c$      & 0.189     & 0.188(1)     \\ \hline
$K^0$       & 0.207$i$  & 0.207(1)$i$  & $K^{*0}$     & 0.224$i$  & 0.224(1)$i$  \\
$D^0$       & 0.454$i$  & 0.454(1)$i$  & $D^{*0}$     & 0.493$i$  & 0.493(1)$i$  \\
$B^0$       & 0.355$i$  & 0.356(1)$i$  & $B^{*0}$     & 0.377$i$  & 0.378(1)$i$  \\
$B_s$       & 0.275$i$  & 0.275(1)$i$  & $B^*_s$      & 0.283$i$  & 0.284(1)$i$  \\ \hline
\end{tabular}
\caption{\label{tab:Charge radius}The charge radii of pseudoscalar/vector flavor-asymmetric mesons, obtained from Eqs.~(\ref{eq.slope},\ref{eq.int}), are presented, with the corresponding charge distributions illustrated in Fig.\,\ref{fig:meson}. The units are fm.}
\end{table}

We then apply the MEM reconstruction to extract $\rho^{h}_{f\bar{g}}(\tilde{r})$ from $F^{h}_{f\bar{g}}(Q^2)$ for the flavor symmetric and asymmetric, pseudoscalar and vector mesons, considering a VMD model with the mass of the vector $h\bar{h}$ meson \cite{Xu:2024vkn}. We thus aim at providing a comprehensive illustration for both charge density and accumulation across space. This is illustrated in Fig.\,\ref{fig:quark_distribution}, which shows the static spatial distribution of valence (anti)quarks in flavor-asymmetric mesons, along with the corresponding cumulative distribution. The examination of results shows heavier quarks more concentrated in space than lighter ones. Indeed, for the $f\bar{g}$ system, as the mass of the $f$ quark increases, the spatial distribution of the $f$ quark contracts significantly, while the distribution of the $g$ quark expands slightly. These patterns are consistent with the expectations from studies based on generalized parton distributions, \emph{e.g.}\,\cite{Raya:2024glv,Almeida-Zamora:2023bqb}. In extreme cases, such as the $B$ meson, the $b$ quark is confined to a radius of 0.2 fm, while that of the $u/d$ quark exceeds 1 fm. This is specially illustrated by Fig.\,\ref{fig:quark}, where results for the flavor-symmetric mesons and VMD model (prior) are also shown for comparison. In addition, the comparison of distributions for vector and pseudoscalar mesons exposes the impact of spin arrangement: spin-aligned systems (vectors) exhibit quark charge densities with more support at slightly larger distances than those for spin-anti-aligned ones (pseudoscalars), suggesting a larger quark-antiquark separation in the former case. This impact tends anyhow to decrease when quarks become heavier. 
  
Interestingly, an immediate consequence of Eqs.\,(\ref{eq:3d},\ref{eq:2d}) is that the 3D and transverse 2D charge radii can be defined both at zero-momentum from the form factor,
\begin{equation}
\label{eq.slope}
  \left\langle r^2\right\rangle=-\left.6 \frac{d}{d Q^2} F\left(Q^2\right)\right|_{Q^2=0},\,   
  \left\langle b^2\right\rangle=-\left.4 \frac{d}{d Q^2} F\left(Q^2\right)\right|_{Q^2=0}\,,
\end{equation}
or from a quadratic distance averaged with the full-domain charge distribution, 
\begin{equation}
\label{eq.int}
  \left\langle r^{2}\right\rangle=4 \pi \int_0^{\infty} d r r^{2} \rho^{3 \mathrm{D}}(r),\, 
  \left\langle b^{2}\right\rangle=2 \pi \int_0^{\infty} d b b^{2} \rho^{2 \mathrm{D}}(b).
\end{equation}
As far as $F(Q^2)$ and $\rho^{3 \mathrm{D}}(r)$ or $\rho^{2 \mathrm{D}}(b)$ are related by either Eq.\,\eqref{eq:3d} or \eqref{eq:2d}, the two definitions are mathematically equivalent if the density profiles decrease fast enough at large distances (see, {\it e.g.}, Ref.\,\cite{Xu:2024vkn}). Tab.\,\ref{tab:Charge radius} collects the charge radii for pseudoscalar and vector mesons from both Eqs.\,(\ref{eq.slope},\ref{eq.int}), 
and shows that the outputs from one or another are strongly consistent, as a validation crosscheck of the reconstructions for all the mesons. 
For charged mesons, it is evident that the charge radius of vector mesons is larger than that of pseudoscalar mesons. However, for neutral mesons, the positive slope of the form factor leads to an imaginary charge radius, as an indication of a non-trivial charge distribution which will be further elaborated.

Then, combining the quark and antiquark densities displayed in Fig.\,\ref{fig:quark_distribution} according to Eq.\,\eqref{eq.charge}, meson's charge distributions are calculated and shown in Fig.\,\ref{fig:meson}. For charged mesons, the constructive superposition of positive densities generates shapes which considerably differ when the flavor asymmetry increases. For example, while $K$ meson's charge distribution is smooth and flat, that of $B$ meson exhibits a sharp peak supplemented by a broad tail, although their radii are nearly the same: $\sqrt{\langle r^2_{K^+} \rangle}=0.608$ fm, $\sqrt{\langle r^2_{B^+} \rangle}=0.619$ fm (see Tab.\,\ref{tab:Charge radius}). In the neutral case, the destructive interference results always in the same qualitative picture: a core formed primarily by the charge of the heavy quark, surrounded by a cloud of charge largely of the light quark. Furthermore, the charge distribution steepens with increasing flavor asymmetry, and the neutral radius $\tilde{r}^0$ (or $\tilde{b}^0$ in 2D), defined as the position where the charge distribution crosses zero, satisfies the following ordering:
\begin{equation}
  \tilde{r}^0_{K^0} > \tilde{r}^0_{D^0} > \tilde{r}^0_{B^0} \gtrsim \tilde{r}^0_{B_s},\   \tilde{r}^0_{K^{*0}} > \tilde{r}^0_{D^{*0}} > \tilde{r}^0_{B^{*0}} \gtrsim \tilde{r}^0_{B^*_s}.
\end{equation}
And again, as a consequence of spin arrangement effects, vector's neutral radii are slightly larger than those of the pseudoscalar case, with a difference which gradually disappears for increasing meson mass.

\begin{figure*}[htbp]
\centering 
\includegraphics[width=0.98\textwidth]{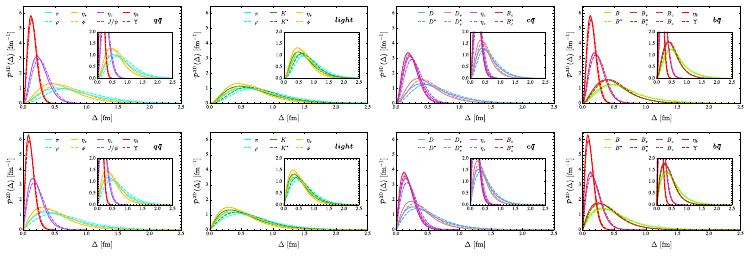}
\caption{\label{fig:delta} The probability density for the distance between quarks and antiquarks in mesons, $\mathcal{P}_{f\bar{g}}(\Delta)$, expressed by Eq.\,\eqref{eq:Deltaprob}.}
\end{figure*} 
 \begin{figure}[htbp]
\centering 
\includegraphics[width=0.45\textwidth]{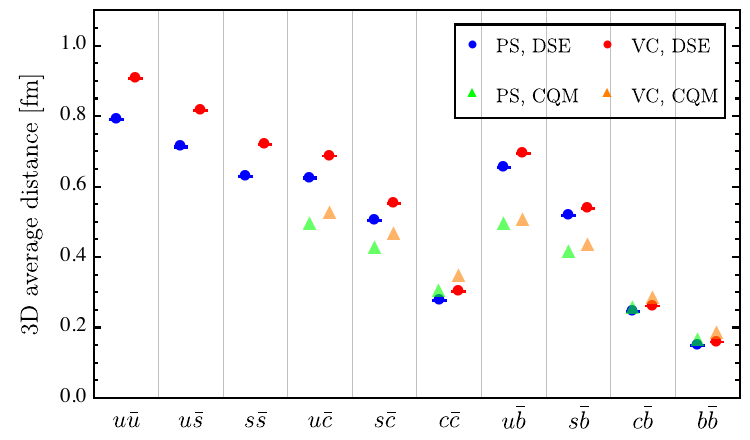}
\caption{\label{fig:comparison} The average distance $\langle \Delta_{f\bar{g}} \rangle$ between quarks and antiquarks in mesons, defined according to Eq.\,\eqref{eq.delta}. More details can be found in Tab.\,\ref{tab:Delta}}.
\end{figure}

An alternative quantity related to meson's spatial extension is the probability density of the distance between the quark and the antiquark forming the meson, expressed as
\begin{equation}\label{eq:Deltaprob}
  \mathcal{P}_{f\bar{g}}(\Delta) = \int d^3\bm{r}_{f} d^3\bm{r}_{\bar{g}}\ \delta(\Delta - |\bm{r}_{f}-\bm{r}_{\bar{g}}|) \rho^{f}_{f\bar{g}}(\bm{r}_{f}) \rho^{\bar{g}}_{f\bar{g}}(\bm{r}_{\bar{g}}),\,
\end{equation}
(or, analogously, in transverse 2D) and normalized to 1 over the full range of positive $\Delta$. Correspondingly, one can define
\begin{equation}
     \langle \Delta_{f\bar{g}} \rangle = \int_0^{\infty}  d\Delta  \, \Delta \, \mathcal{P}_{f\bar{g}}(\Delta)  \label{eq.delta}.
\end{equation}
as the averaged distance for the quark-antiquark pair, which is an expression of the meson size, remaining always positive definite and exposing better the effects of the spin arrangement. This probability density for all flavor-asymmetric compared to the flavor-symmetric mesons is numerically evaluated\,\footnote{It is well known that numerical integrals with a $\delta$-function are often challenging. However, the $\delta$-function in Eq.\,\eqref{eq.delta} can be removed when integrating out one of the radii by performing a coordinate transformation} and displayed in Fig.\,\ref{fig:delta}, while the averaged quark-antiquark distance is collected in Tab.\,\ref{tab:Delta} and displayed in Fig.\,\ref{fig:comparison}. For all mesons, the probability density presents a peak which, in $q\bar{q}$, $b\bar{q}$ and $c\bar{q}$ systems ($q=u,s,c,b$), shifts gradually towards low distances as the current-quark mass $m_q$ increases, while the width shrinks. The effect is negligible when only light quarks ($u$ and $s$) are involved. The table shows that, in the flavor-symmetric case, $\langle \Delta_{q\bar{q}} \rangle$ gradually decreases as the current-quark mass $m_q$ increases, with $\langle\Delta_{u\bar{u}}\rangle/ \langle\Delta_{b\bar{b}}\rangle \sim 5$, while $m_b \approx 1000\ m_u$. Furthermore, the following hierarchies can be seen to appear
\begin{subequations}
\label{eq:trend}
\begin{align}
  &\langle \Delta_{B} \rangle >  \langle \Delta_{B_s} \rangle >  \langle \Delta_{B_c} \rangle >  \langle \Delta_{\eta_b} \rangle,\  \langle \Delta_{B^*} \rangle >  \langle \Delta_{B^*_s} \rangle >  \langle \Delta_{B^*_c} \rangle >  \langle \Delta_{\Upsilon} \rangle,  \\
    &\langle \Delta_{D} \rangle >  \langle \Delta_{D_s} \rangle >  \langle \Delta_{\eta_c} \rangle >  \langle \Delta_{B_c} \rangle,\ \langle \Delta_{D^*} \rangle >  \langle \Delta_{D^*_s} \rangle >  \langle \Delta_{J/\psi} \rangle >  \langle \Delta_{B^*_c} \rangle.
\end{align}
\end{subequations}
While, when comparing mesons containing the same flavor structure, one is left with: $\langle \Delta_{\text{VC}} \rangle> \langle\Delta_{\text{PS}}\rangle$, confirming the spin effect. This latter, can be quantified by defining 
\begin{equation}\label{eq:lambda}
   \lambda=(\langle\Delta_{\text{VC}}\rangle-\langle\Delta_{\text{PS}}\rangle)/\langle\Delta_{\text{PS}}\rangle,
\end{equation} 
which is determined to be roughly $15\%$ for light mesons and, respectively, $10\%$ and $5\%$ when quarks $c$ and $b$ are involved. This confirms that the spin effect gradually weakens when the quark masses increase and the meson approaches the non-relativistic limit. On the other hand, the transverse 2D inter-distances obey the same qualitative pattern but become reduced by a factor of 1.273(1). 

\begin{table}[htbp]
\centering
\begin{tabular}{c c c | c c c}
\hline
 & $\langle \Delta \rangle^{\text{2D}}$ & $\langle \Delta \rangle^{\text{3D}}$ &
 & $\langle \Delta \rangle^{\text{2D}}$ & $\langle \Delta \rangle^{\text{3D}}$ \\
\hline
$\pi$     & 0.621(2) & 0.791(2) & $\rho$     & 0.712(1) & 0.907(1) \\
$K$       & 0.560(2) & 0.713(3) & $K^*$      & 0.642(1) & 0.817(1) \\
$\eta_s$  & 0.493(1) & 0.628(1) & $\phi$     & 0.565(1) & 0.720(1) \\
$D$       & 0.490(2) & 0.624(2) & $D^*$      & 0.540(1) & 0.687(1) \\
$D_s$     & 0.394(1) & 0.503(1) & $D^*_s$    & 0.433(1) & 0.552(1) \\
$\eta_c$  & 0.217(1) & 0.277(1) & $J/\psi$   & 0.238(1) & 0.303(1) \\
$B$       & 0.514(1) & 0.654(1) & $B^*$      & 0.544(1) & 0.693(1) \\
$B_s$     & 0.407(1) & 0.518(1) & $B^*_s$    & 0.422(1) & 0.537(1) \\
$B_c$     & 0.192(1) & 0.245(1) & $B^*_c$    & 0.205(1) & 0.261(1) \\
$\eta_b$  & 0.117(1) & 0.149(1) & $\Upsilon$ & 0.125(1) & 0.159(1) \\
\hline
\end{tabular}
\caption{The average distance $\langle \Delta_{f\bar{g}} \rangle$ between quarks and antiquarks in pseudoscalar and vector mesons. For the sake of comparison, one can refer to results from CQM \cite{Deng:2020iqw} for 3D inter-distances: $D$($D^*$)=0.50,0.53; $D_s$($D^*_s$)=0.43,0.47; $\eta_c$($J/\psi$)=0.31,0.35; $B$($B^*$)=0.50,0.51; $B_s$($B^*_s$)=0.42,0.44; $B_c$($B^*_c$)=0.26,0.29; $\eta_b$($\Upsilon$)=0.17,0.19. The units are fm. An intuitive comparison is illustrated in Fig.\,\ref{fig:comparison}.}
\label{tab:Delta}
\end{table}

Strikingly, the very same trend shown by Eqs.\,\eqref{eq:trend} and main qualitative features related to average distances have been also recently reported from studies based on constituent quark models (CQM)\,\cite{Deng:2020iqw} and the radii inferred from the contact interaction model\,\cite{Hernandez-Pinto:2023yin,Hernandez-Pinto:2024kwg} (see Tab.\,\ref{tab:Delta} and Fig.\,\ref{fig:comparison}). Particularly, in the heavy quarkonia sector, quantitative agreement can be also claimed, while discrepancies when light quarks are involved can be attributed to the potential model approach of CQM.      

\begin{figure*}[htbp]
\centering 
\includegraphics[width=0.98\textwidth]{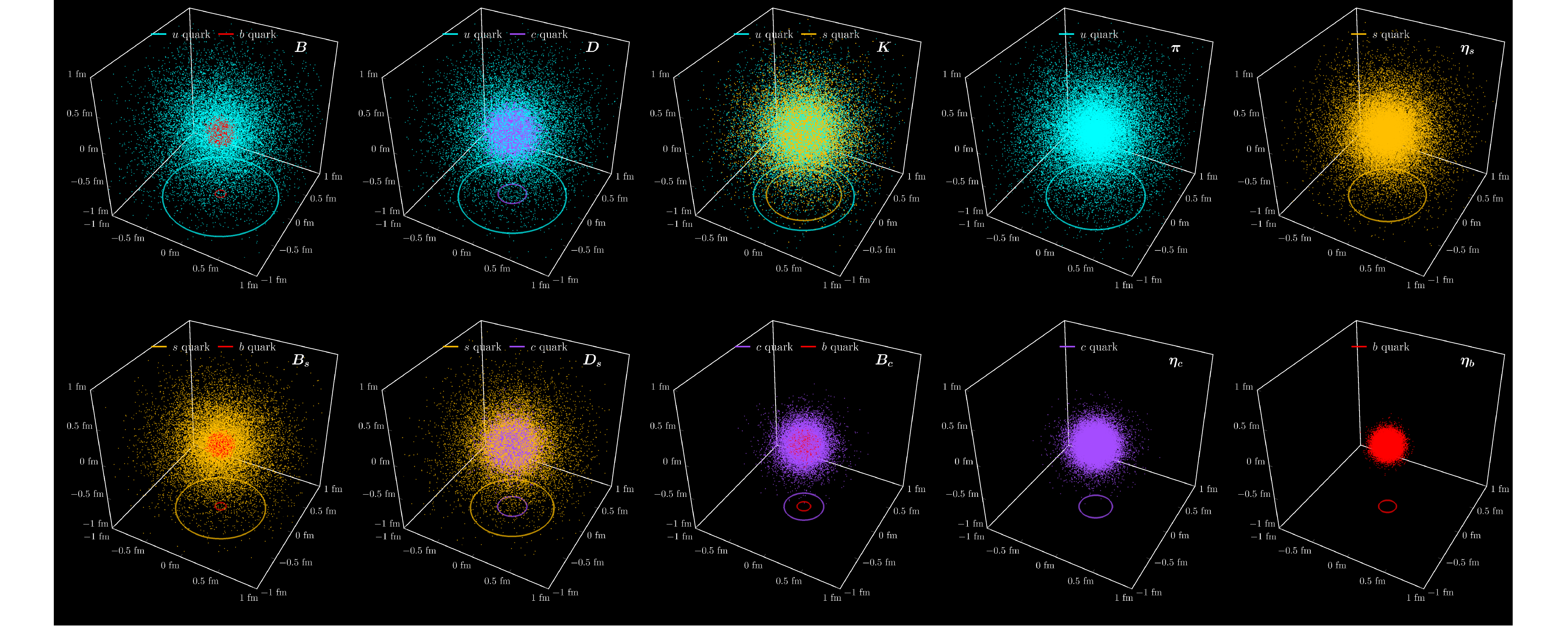}
\caption{\label{fig:ps_sample} Under electromagnetic probe, the 3D spatial static distribution of valence (anti)quark in pseudoscalar mesons, which are constructed based on $2 \times 20000$ sample points. The circles at the bottom denote the corresponding charge radius. }
\end{figure*}

Finally, aiming at a pictorial representation for the charge structure of the mesons, Fig.\, \ref{fig:ps_sample} has been produced. There, we generate $20000$ sample points for each quark location, with random angular distribution and a radial density given by the obtained 3D quark distributions, and display them in 3D spherical plots. Clearly, a $B$ meson exhibits a dense core enveloped by a cloud of light quarks, and the latter collapses into the former as flavor asymmetry is reduced. It is also apparent that the size of cores and clouds diminishes with increasing quark masses. We have explicitly shown the pseudoscalar case only. However, the picture for vector mesons remains unchanged as expected from the density profiles shown in Figs.\,\ref{fig:quark_distribution} and \ref{fig:quark}, only with slightly larger sizes for clouds and cores of lighter quarks. 

\section{Summary and discussions}
\label{sec:4} 

We have capitalized on available results for the electromagnetic form factors of flavor-symmetric and -asymmetric, ground-state pseudoscalar and vector mesons, which were obtained within the DSEs/BSEs framework, to derive their charge distributions and identify the individual contributions from different quark flavors. One of our primary aims is to illustrate the reliability of the previously proposed and optimized Maximum Entropy Method for reconstructing these distributions. The resulting charge distributions can thus provide a physically intuitive picture of strongly interacting bound states.

The analysis has been carried out in both 3D and transverse 2D cases. While both display broadly similar qualitative distribution shapes and overall structures, the 2D transverse distributions are quantitatively more compact. Specifically, independent quark distributions exhibit a characteristic profile featuring a pronounced central peak and a broad tail. When these features are combined into the complete meson distribution, the picture that emerges is one of a heavy-quark charge core surrounded by a lighter-quark charge cloud. As the flavor asymmetry begins to decrease, the core-cloud structure becomes less conspicuous. These two regions gradually merge and the overall charge distribution becomes flatter.

In addition, we propose the average distance between the quark and the antiquark forming the pair as an insightful scale characterizing the size of mesons and the typical range of quark motion within them. A clear hierarchy is observed: for quarkonia, the heavier the quark, the more compressed the meson (e.g., the average distance between the valence quark and antiquark in the $u\bar{u}$ system is approximately five times larger than in the $b\bar{b}$ system). In flavor-asymmetric mesons, the average distance increases with growing asymmetry. A comparison between pseudoscalar and vector mesons of the same flavor composition reveals a noticeable spin effect: vector mesons exhibit distances that are 5–15\% larger than their pseudoscalar counterparts. This spin effect diminishes with increasing quark masses, consistent with a transition from relativistic to non-relativistic regimes. Notably, this pattern aligns with findings from earlier studies based on constituent quark models. 

The same reconstruction procedure herein used can be explored in the future to unveil structural properties under electromagnetic probe of other hadron states, including multiquark systems. 

%
%\section*{Acknowledgments}
\medskip
\noindent\textbf{Acknowledgments}.
This work has been partially funded by Ministerio Espa\~nol de Ciencia e Innovaci\'on under grant Nos. PID2019-107844GB-C22 and PID2022-140440NB-C22; Junta de Andaluc\'ia under contract Nos. Operativo FEDER Andaluc\'ia 2014-2020 UHU-1264517, P18-FR-5057 and also PAIDI FQM-370. The authors acknowledge, too, the use of the computer facilities of C3UPO at the Universidad Pablo de Olavide, de Sevilla.

\medskip
\noindent\textbf{Data Availability Statement}. This manuscript has no associated data or the data will not be deposited. [Authors' comment: All information necessary to reproduce the results described herein is contained in the material presented above.]

\medskip
\noindent\textbf{Declaration of Competing Interest}.
The authors declare that they have no known competing financial interests or personal relationships that could have appeared to influence the work reported in this paper.

\bibliographystyle{model1a-num-names}
\bibliography{ref.bib}

\end{document}